\documentclass[iop]{emulateapj}

\usepackage{graphicx}
\usepackage{latexsym}
\usepackage{amsfonts,amsmath,amssymb}
\usepackage{url}
\usepackage[utf8]{inputenc}
\usepackage{fancyref}
\usepackage{hyperref}
\hypersetup{colorlinks=false,pdfborder={0 0 0},}

\usepackage{natbib}

\newcommand{\na}{New Astronomy}

\shorttitle{Filamentary bundles in molecular clouds}
\shortauthors{Moeckel \& Burkert}

\begin{document}

\title{The formation of filamentary bundles in turbulent molecular clouds}

\author{Nickolas Moeckel}
\affil{University Observatory Munich, Scheinerstrasse 1, D-81679 Munich, Germany}
\email{nickolas1@gmail.com}
 
\author{Andreas Burkert}
\affil{University Observatory Munich, Scheinerstrasse 1, D-81679 Munich, Germany}
\affil{Max-Planck-Institute for Extraterrestrial Physics, Giessenbachstrasse 1, D-85758 Garching, Germany}
\email{burkert@usm.uni-muenchen.de}

\begin{abstract}
The classical picture of a star-forming filament is a near-equilibrium
structure, with collapse dependent on its gravitational criticality.
Recent observations have complicated this picture, revealing filaments
as a mess of apparently interacting subfilaments, with transsonic
internal velocity dispersions and mildly supersonic intra-subfilament
dispersions. How structures like this form is unresolved. Here we study
the velocity structure of filamentary regions in a simulation of a
turbulent molecular cloud. We present two main findings: first, the
observed complex velocity features in filaments arise naturally in self
gravitating hydrodynamic simulations of turbulent clouds without the
need for magnetic or other effects. Second, a region that is filamentary
only in projection and is in fact made of spatially distinct features
can displays these same velocity characteristics. The fact that these
disjoint structures can masquerade as coherent filaments in both
projection and velocity diagnostics highlights the need to continue
developing sophisticated filamentary analysis techniques for star
formation observations.

\end{abstract}

\bibliographystyle{apj}

\section{Introduction} 
Observations with the Herschel satellite have revealed that the backbone of molecular clouds is a compex network of connecting and interacting filaments \citep[e.g.][]{2010A&A...518L.102A,2010A&A...518L.100M,2011A&A...529L...6A,2012A&A...540L..11S}. The dynamics of this web of dense molecular gas not only determines the evolution and stability of molecular clouds, but also regulates their condensation into stars. Molecular cloud cores and single low-mass stars are almost always found in filaments, often aligned like pearls on a string \citep{2002ApJ...578..914H,2008ApJ...672..410L,2010A&A...518L.102A}. This can be interpreted as a result of the gravitational instability of a supercritical filamentary section \citep{1997ApJ...480..681I,2011A&A...533A..34H}. Where filaments intersect, more massive hubs form \citep{2009ApJ...700.1609M,2013ApJ...764..140M} that later on could become the progenitors of star clusters.

\citet{2010ApJ...724..687L} \citep[see also e.g.][]{2007ApJ...666..982E} demonstrated that the fraction of gas in the dense molecular web is of order 10 percent of the total molecular mass. Interestingly, the mass fraction of protostars to dense ($n > 10^4$ cm$^{-3}$) molecular gas is also a constant of the same order. \citet{2013ApJ...773...48B} showed that this requires new filamentary segments to continuously form from the diffuse intra-filament medium on a gravitational collapse timescale.

Classically, gas filaments have been treated as cylinders of gas in hydrostatic equilibrium \citep{1964ApJ...140.1529O,2012A&A...542A..77F,2013A&A...558A..27R}. It is not clear, however, how such quiescent structures could form in the turbulent environment of a molecular cloud \citep[e.g.][]{2010A&A...520A..17K,2013ApJ...769..115H}. In addition, millimeter line studies indicate that filaments have intrinsic, super-thermal linewidths \citep{2013A&A...553A.119A}.

More recently, observations by \citet{2013A&A...554A..55H} revealed that filaments are often compact bundles of thin spaghetti-like subfilaments. They presented observations of a prominent filamentary feature in Taurus, dominated by the L1495 cloud \citep{1962ApJS....7....1L} and several dark patches \citep{1927cdos.book.....B}. They observed the region in the moderate density tracer C$^{18}$O, obtaining spectra along a $\sim 10$ pc length of the region. Analysing the richest $\sim 3$ pc section of the filament in the resultant position--position--velocity space, they found the intriguing result that the gas along the ridge is organized in velocity-coherent filamentary structures, with typical lengths $\sim0.5$ pc. Each filament is internally subsonic or transsonic, though the collection of filaments is characterised by a mildly supersonic interfilamentary dispersion of $\sim 0.5$ km s$^{-1}$. They describe the collection of velocity detections as "elongated groups that lie at different `heights' (velocities) and present smooth and often oscillatory patterns".
Multiple velocity components along a single line of sight have also been observed in Serpens South \citep{2013ApJ...778...34T}; this feature may therefore be common for many young star forming sites.

The complexity of Hacar et al.'s position--position--velocity data, exemplified by their figure 9, invites theoretical and numerical exploration. In particular the apparent organization of filaments into bundles tends to bring to mind magnetic fields or other relatively complex physics.
In this paper we explore whether this velocity signal is present in simulated molecular clouds with a more minimal set of physics, including only gravity and hydrodynamic forces acting on the initial turbulence.

\begin{figure*}
\begin{center}
\includegraphics[width=2\columnwidth]{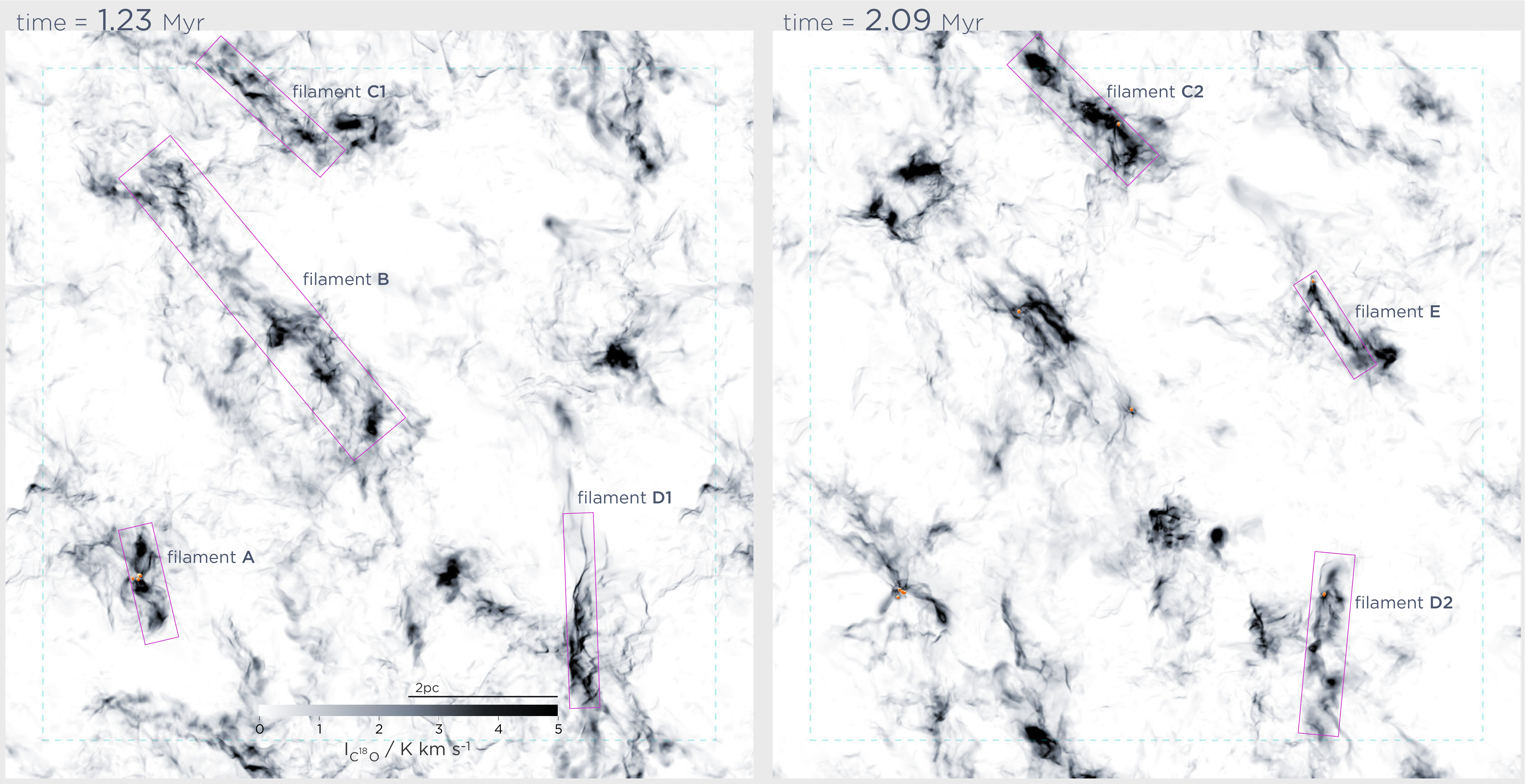}
\caption{\label{finderimage}
Surface density projections through the simulated volume at two times. The dashed cyan lines mark the boundary of the periodic simulation domain. We calculated the surfaced density using only the gas between $10^3 < n/{\rm cm}^{-3} < 10^{4.5}$, and converted this to an approximation of an optically thin C$^{18}$O observation. The magenta boxes delineate the filaments we focus on in this paper. Sink particles are shown as orange dots.}
\end{center}
\end{figure*}

\section{The simulation}
We simulated a 10 pc periodic cube of a molecular cloud, beginning from fully developed isothermal turbulent initial conditions. We generated the turbulent initial conditions using version 4.2 of the ATHENA code \citep{2005JCoPh.205..509G,2008JCoPh.227.4123G,2008ApJS..178..137S,2009NewA...14..139S}. The turbulent driving was technically similar to that of \citet{2009ApJ...691.1092L}. Briefly, on a $1024^{3}$ grid with domain length 1 we applied divergence-free velocity perturbations at every timestep to an initially uniform medium. The perturbations had a Gaussian random distribution with a Fourier power spectrum $\left | d {\bf v}_{k}^{2} \right | \propto k^{-2}$,
for wavenumbers $2 < k/2 \pi < 4$. Similarly to, e.g., \citet{Federrath_2013}, turbulence on smaller length scales was generated self-consistently from the large scale driving. The driving continued until the box reached a saturated state at a Mach number $\mathcal{M}$ of roughly 8. During this driving stage we did not include gravitational forces.

\begin{figure*}
\begin{center}
\includegraphics[width=2\columnwidth]{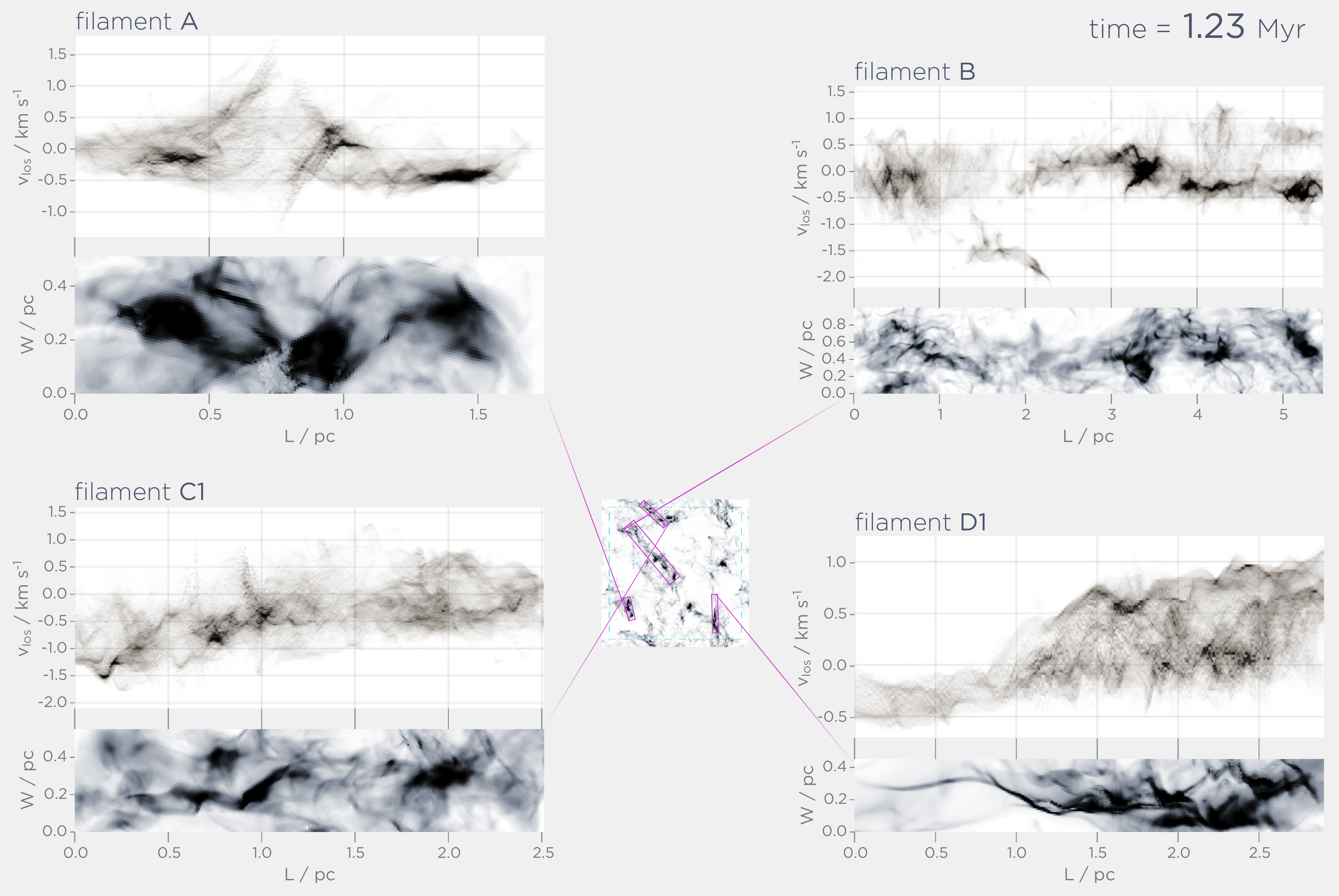}
\caption{\label{filvels1}
The filaments from the snapshot at 1.23 Myr, shown in their native coordinate system, along with line of sight velocity observations along the $L$ dimension. These density weighted line of sight velocities are summed along the $W$ dimension, i.e. they are the $L$--$v$ plane of the $L$--$W$--$v$ position--position--velocity cube.}
\end{center}
\end{figure*}

At this point we scaled the box to a physical size $S = 10$ pc, mean number density $n = 100$ cm$^{-3}$ at mean molecular weight $\mu=2.33$ (giving a total gas mass $\sim5700$ M$_{\sun}$), and a constant sound speed $c_s$ = 0.2 km s$^{-1}$. After turning off the forcing and turning on self gravity, we turned the simulation over to the adaptive mesh refinement (AMR) code RAMSES \citep{2002A&A...385..337T}. The $1024^{3}$ base grid was maintained, along with 2 steps of adaptive refinement triggered when the local Jeans length became shorter than 4 grid cells \citep{1997ApJ...489L.179T}. 

Regions collapsing beyond this point were replaced by sink particles, using the sink implementation described in \citet{2010MNRAS.409..985D}, which largely follows the implementation of \citet{2004ApJ...611..399K}. The salient points of this implementation are that regions exceeding the Truelove density limit on the finest level of refinement and that are collapsing along all directions are replaced by sink particles. The sinks accrete gas in a momentum conserving fashion from a region 4 cells in radius ($\sim0.01$ pc) at the local Bondi-Hoyle rate in that region. While the sinks are addressed in this paper, we can approximate the integrated star (rather, sink) formation efficiency as the fraction of the total mass in sinks at some time.

We integrated the box through $\sim2.1$ Myr of self gravitating evolution. Following \citet{2007ApJ...665..416K}'s definition of the turbulent turnover time in a box , $T_{turb} = S/(2 c_s \mathcal{M})$, we have $T_{turb} \sim 3$ Myr. The free-fall time  $T_{ff} = [3\pi / (32 G \rho)]^{1/2}$, at the mean density of the simulation, is also roughly 3 Myr. In this paper we focus on moderately dense gas, with $10^3 < n/{\rm cm}^{-3} < 10^{4.5}$; at these densities we have $0.2 \lesssim T_{ff} / \rm{Myr} \lesssim 1.1$. After 1 Myr the global structure of the simulated box is thus still dominated by the initial turbulence, while the denser gas has had time to become organized by self gravity.

\section{Filament selection and analysis}
After 1.23 Myr and 2.09 Myr of self gravitating evolution, we selected by eye filamentary features in one projection of the cloud, shown in Figure \ref{finderimage}.  These images were constructed by calculating the surface density of gas in the range $10^3 < n/{\rm cm}^{-3} < 10^{4.5}$. This is roughly the density range traced by C$^{18}$O observations; we then converted the surface density to integrated line intensities. This conversion, using equation 2 of \citet{2001ApJ...551..687M}, is only meant to give a low-order approximation to an observation. This approximate `observation' of our data yields an idea of what this simulation might look like in the motivating observations of \citet{2013A&A...554A..55H}. At 1.23 Myr we chose one feature which, observationally speaking, would probably be deemed an obvious filament (filament D1), two that might be considered a ridge of star formation activity consisting of distinct dense sites linked by more tenuous emission (filaments B and C1), and one region that has entered a collapse phase (filament A). We stress that this is not intended to be a comprehensive study of all the filaments in this projection, but rather an exploratory look at a few of the more prominent features. At this point 34 M$_{\sun}$ of gas is in sinks in filament A, yielding a global integrated sink formation efficiency of $\sim 0.6\%$.

At 2.09 Myr we selected three filamentary features. Two are evolved versions of the features at 1.23 Myr (filaments C2 and D2), while filament E formed in the intervening time. At this point there are more sinks scattered through the box, including at least one in each of the filaments. The total sink mass is 92 M$_{\sun}$, i.e. a global integrated sink formation efficiency $\sim1.6\%$\footnote{We note in passing that this efficiency is a factor of 1.8 lower than the most equivalent simulation of \citet{2012ApJ...761..156F}, within the expected scatter resulting from different turbulent seeds, box sizes, and densities.}. We chose to analyse these early times, before the gas has begun converting to sinks in earnest, to approximate the evolutionary state of Hacar et al.'s observations. The masses and dimensions of the sinks are summarized in Table \ref{filtable}.

\begin{figure*}
\begin{center}
\includegraphics[width=2\columnwidth]{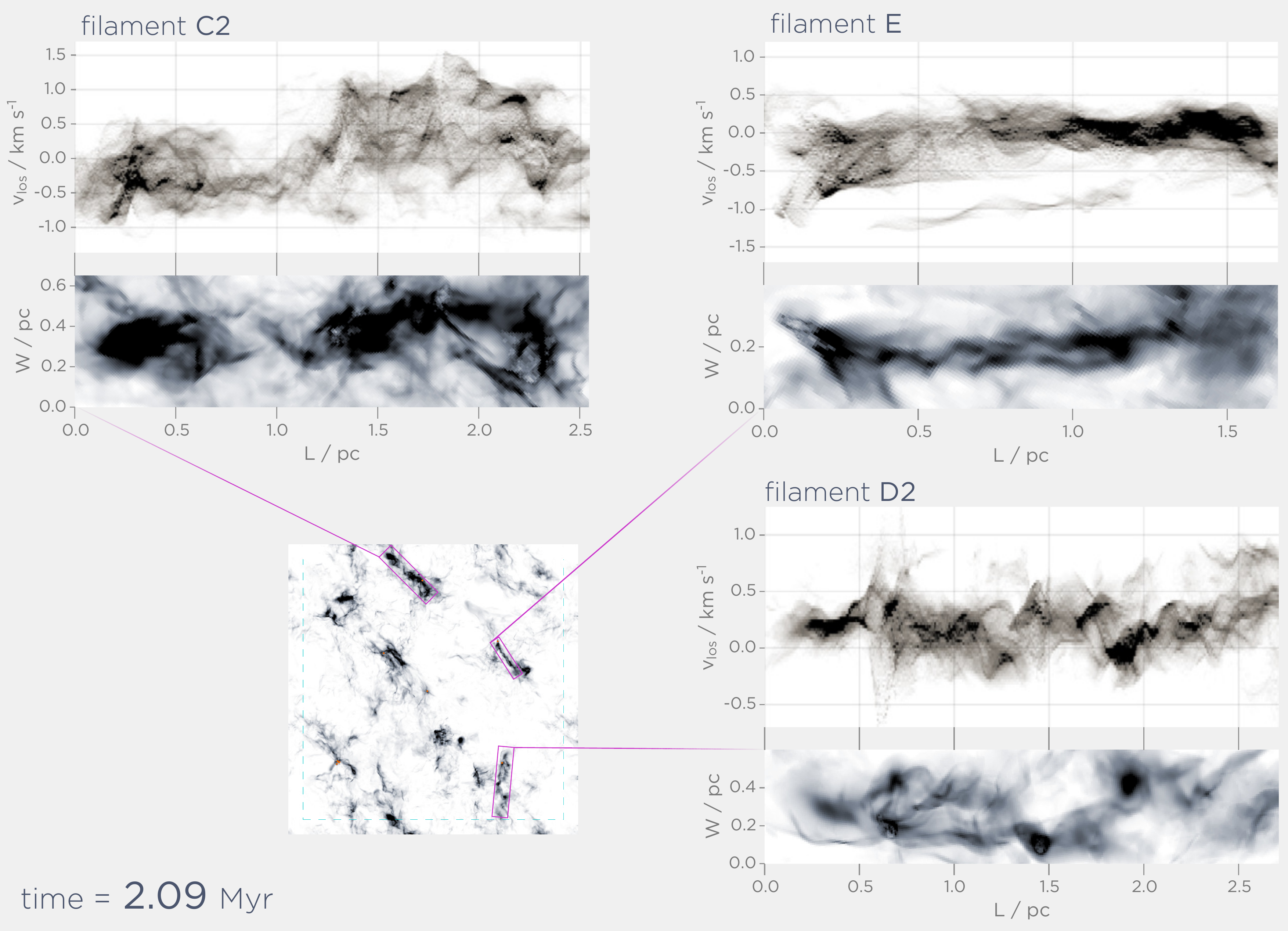}
\caption{\label{filvels2}
The filaments from the snapshot at 2.09 Myr, shown in their native coordinate system, along with line of sight velocity observations along the $L$ dimension. These density weighted line of sight velocities are summed along the $W$ dimension, i.e. they are the $L$--$v$ plane of the $L$--$W$--$v$ position--position--velocity cube.}
\end{center}
\end{figure*}

\subsection{Line-of-sight velocity structure}
The rectangles surrounding the selected filaments define a coordinate system with axes aligned along the long axis or length $L$ of the filament, and the width $W$. The depth $D$ is the same axis that the surface density is projected along. We stepped along the $L$ and $W$ coordinates with a stepsize equal to the simulation's base resolution, and calculated at each point the density-weighted line-of-sight velocity integrated along the $D$ axis, using the same density range used to calculate the surface density. We binned these spectra into histograms with bin width 0.025 km s$^{-1}$.

In Figures \ref{filvels1} and \ref{filvels2} we show this data above the blowups of each filament. We summed the histograms along the $W$ dimension, yielding the total velocity distribution along slices perpendicular to the filament's long axis. The character of the velocity distributions are qualitatively similar to the observations of \citet{2013A&A...554A..55H}; overlapping individual features with subsonic widths, combined into features with mildly supersonic dispersions. Note that we plot the raw density weighted velocity structure, rather than the centroids of line fits like Hacar et al present. This accounts for the low intensity background in the velocity plots. Nonetheless, arcs of higher intensity signal are clearly seen in our data.

There are a few features in individual filaments worth mentioning. Filaments C1 and D1 show large scale gradients of approximately 0.5 km s$^{-1}$ pc$^{-1}$ on top of which the $\sim 0.5$ km s$^{-1}$ dispersion is overlaid. The small scale feature's velocity gradients are larger than the large scales, consistent with the observations. Filament A's sink particles are clustered around $L = 0.75$ pc; the inflow of material onto this small protocluster is clearly visible in the velocity plot, while farther away from the clustering the intertwining filaments in position--velocity space again show up. This same signal appears in filament D2, with about 6 M$_{\sun}$ of sinks at $L=0.6$ pc. Filament B contains outlying material at a velocity of about -1.5 km s$^{-1}$, quite distinct from the rest of the signal at mean velocities around 0 to -0.5 km s$^{-1}$, a hint that some material in the ridge is only associated with the rest in projection. To further explore the coherence of the filaments, we turn to the 3D spatial extent of the selected regions at 1.23 Myr.

\subsection{The filaments' third spatial dimension}
\begin{figure*}
\begin{center}
\includegraphics[width=2\columnwidth]{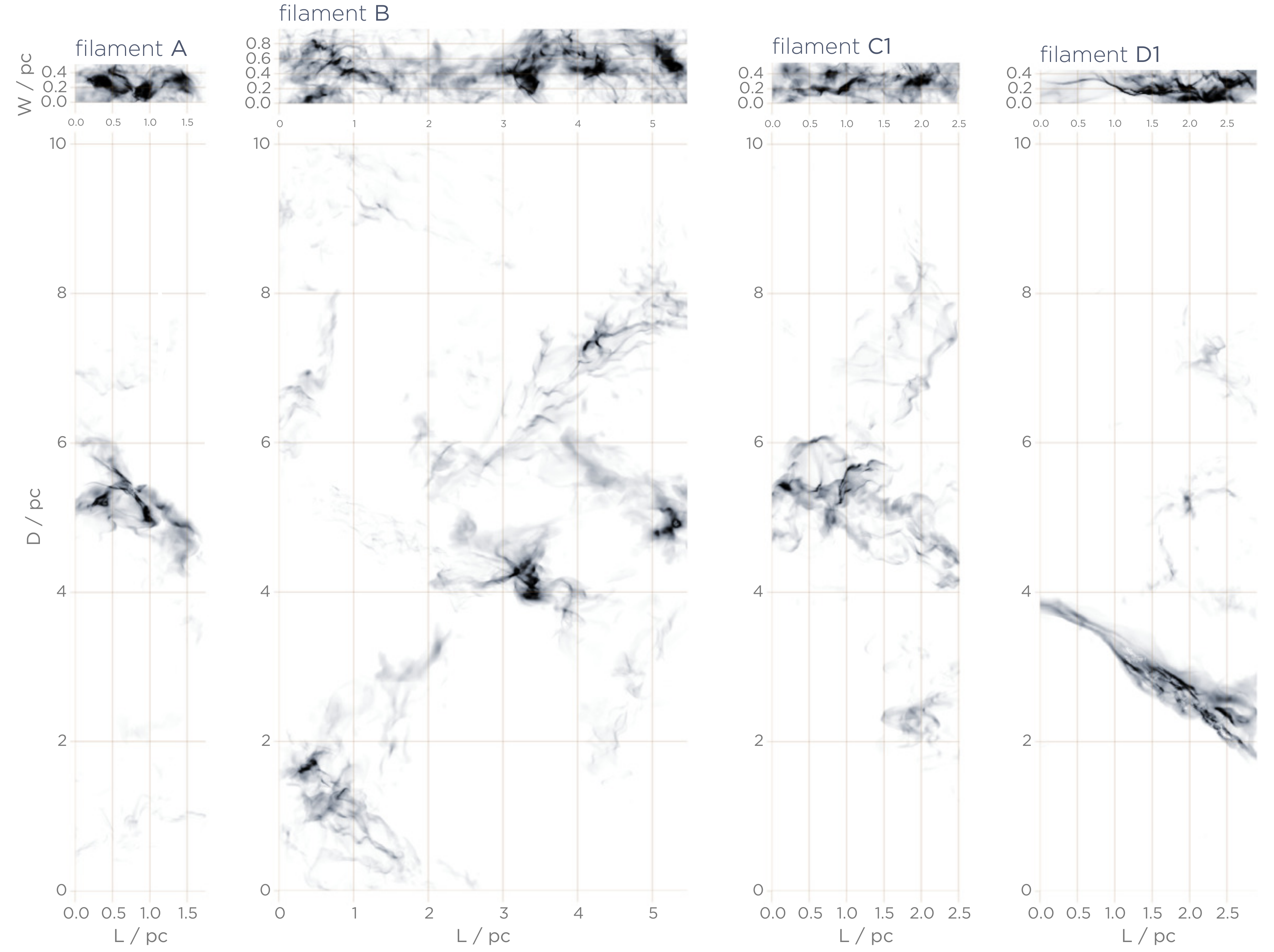}
\caption{\label{filaments3D}
The third dimension of the filamentary features from 1.23 Myr. The projections in the $L$--$D$ plane show only the material that contributes to the surface density in the regions marked by the magenta boxes in the left panel of Figure \ref{finderimage}, shown again here in the top panels.}
\end{center}
\end{figure*}

In Figure \ref{filaments3D} we show the $L$--$D$ projections along with the $L$--$W$ projections from the left panel of Figure \ref{finderimage}. Filaments A, C1, and D1 are revealed as spatially coherent entities along the projected axis, with most of the material contained in a single connected distribution. Filament D1, in particular, is predominantly a 1D structure. Filament A, as the first site of sink particle formation in the simulated volume, is unsurprisingly compact in all three dimensions. Filament B, in contrast, consists of distinct dense clumps, no closer to each other than about 2 pc, that are joined only in projection by more tenuous emission. 

It is worth noting that there is nothing in the line-of-sight velocity information that obviously distinguishes filament B from C1 or D1. With the exception of the outlying material at $L\sim 1.5$ to $2.25$ pc, and $v_{los}\sim-1.5$ to $-2$ km s$^{-1}$ (which is the clump at $D\sim 2.5$ to $3.5$ pc in Figure \ref{filaments3D}), the rest of the material appears to consist of distinct ribbons intertwining in position--velocity space, with some large scale gradients and a mildly supersonic dispersion.

\begin{table}
\begin{center}
\begin{tabular}{lcccccc}
\hline
\vspace{-0.15cm}
		& $M_{C^{18}O}$	& $M_{gas}$	& $M_{\star}$	& 		& $L$	& $W$\\ \vspace{-0.15cm}
filament	&				&			&			& $sfe$	&		&\\ 
      		& (M$_{\sun}$) 		& (M$_{\sun}$) & (M$_{\sun}$) &  		&  (pc) 	& (pc) \\   
\hline      
A           & 54.3      & 114     & 0   & 0 & 1.7   & 0.50   \\            
B       & 194   & 549  & 0 & 0 & 5.5 & 1.0   \\
C1        & 69.8  &  160  & 34.1 & 0.17 & 2.5 & 0.55  \\
D1     & 67.9  & 134     & 0 &  0 & 2.9  & 0.45    \\
C2       & 130  & 246     & 4.95 & 0.02 & 2.5  & 0.65    \\    
D2      & 66.9  & 161     &7.41 &  0.044 & 2.7  & 0.60    \\
E      & 36.0  & 79.1     & 1.21 &  0.015 & 1.7  & 0.40    \\
\hline
\end{tabular}
\caption{The masses and dimensions of the filaments as defined by the boxes in Figure \ref{finderimage}. $M_{C^{18}O}$ is the moderate density mass contributing to that image;  $M_{gas}$ is the total gass mass; $M_{\star}$ is the mass in sinks; $sfe$ is the global sink formation efficiency in each filament, $M_{\star} / (M_{\star} + M_{gas})$.}
\label{filtable}
\end{center}
\end{table}

\section{Concluding discussion}
We simulated a 10pc region of a turbulent molecular cloud, allowing it to evolve under self gravity for 1.25 Myr. While the global structure of the region is still dominated by its turbulent structure, the denser gas has had time to become gravitationally organized. We treated the simulation as an observer might observe the sky; converting the density to an approximate line intensity, selecting filamentary regions for further study, and analysing the line-of-sight velocity information in these regions.

Our first main finding is that velocity characteristics very similar to those observed by \citet{2013A&A...554A..55H} form naturally in such a turbulent setup. Individually bound subfilaments display approximately sonic or subsonic dispersions, while the agglomerations that make up the larger filaments have transsonic to mildly supersonic relative motions. While we will further study the detailed evolution of these structures in future work, we speculate that this velocity structure is a relic of the supersonic turbulence that is generally taken as the initial conditions of star formation. The substructured (both in space and velocity) filaments appear without the need for magnetic fields, which are the physical mechanism that immediately spring to mind when considering a filament composed of a bundle of subfilaments.

Secondly, the same velocity structures that appear in spatially coherent filaments can show up when observing a faux-filament. Based on the line-of-sight velocity information, there is not a clear case to be made that our filament B (particularly the section with $L > 2$pc; see Figure \ref{filvels1}) is different from filaments C1 or D1. Examining the third dimension reveals it to in fact be composed of widely separated dense regions (Figure \ref{filaments3D}). Apparent velocity coherence similar to that seen in our most linear structures is evidently not enough to diagnose a true filament. The presence of this imposter filament in our data, not merely in projected density but also in the line-of-sight velocity, highlights the need for the continued development of sophisticated filament diagnostics, using both simulation and observation and both spatial and velocity data, in order to interpret not only existing observations of star forming filaments but also the imminent onslaught of data from ALMA.

\acknowledgements
This project has been funded by the priority program 1573 "Physics of the interstellar medium" of the German Science Foundation. Using {\tt yt} \citep{2011ApJS..192....9T}\footnote{{\tt www.yt-project.org}} made analysing the simulated data a great deal more efficient and painless than it otherwise would have been. NM sincerely thanks the development team for their rapid resolution of problems and squashing of bugs. In parts of this work we made use of Astropy\footnote{{\tt www.astropy.org}}, a community-developed core Python package for Astronomy \citep{2013A&A...558A..33A}. Our thanks as well to Alvaro Hacar and Jo\~ao Alves for helpful comments and discussions.

\end{document}